\documentclass[twocolumn]{autart}    

\usepackage{graphicx}          

\usepackage{amsmath,amssymb}                     
\usepackage{bm}          

\begin{document}

\begin{frontmatter}

\title{Value Function in Maximum Hands-off Control\thanksref{footnoteinfo}} 

\thanks[footnoteinfo]{This paper was not presented at any IFAC 
meeting. Corresponding author M.~Nagahara.}

\author[Kyoto]{Takuya Ikeda}\ead{ikeda.t@acs.i.kyoto-u.ac.jp},    
\author[Kyoto]{Masaaki Nagahara}\ead{nagahara@ieee.org}

\address[Kyoto]{Graduate School of Informatics, Kyoto University, Kyoto 606-8501, Japan}  
          
\begin{keyword}
Optimal control, continuity, bang-bang control,
discontinuous control, linear systems, minimum-time control.
\end{keyword}

\begin{abstract}
In this brief paper, we study the value function in maximum hands-off control.
Maximum hands-off control, also known as sparse control,
is the $L^0$-optimal control among the admissible controls.
Although the $L^0$ measure is discontinuous and non-convex,
we prove that the value function, or the minimum $L^0$ norm of the control,
is a continuous and strictly convex function of the initial state in the reachable set,
under an assumption on the controlled plant model.
This property is important, in particular, for discussing the sensitivity of the optimality
against uncertainties in the initial state, and also for
investigating the stability by using the value function as a Lyapunov function
in model predictive control.
\end{abstract}

\end{frontmatter}

\section{Introduction}
\label{sec:introduction}

Optimal control is widely used in recent industrial products
not just for achieving the best performance but 
for reducing the control effort.
For example, the classical LQR (Linear Quadratic Regulator) control
gives a way to consider the tradeoff between performance and control-effort reduction
by using weighting functions on the states and the control inputs
with the $L^2$ norm (i.e. the energy) \cite{AndMoo}.

Recently, a novel control method, called {\em maximum hands-off control},
that maximizes the time duration
in which the control is exactly zero among the admissible controls
\cite{NagQueNes13,NagQueNes14b}.
An example of hands-off control is a stop-start system in
automobiles, in which an automobile automatically shuts
down the engine (i.e. zero control) to avoid it idling for long periods of time,
and also to reduce CO or CO2 emissions as well as fuel
consumption.
Therefore, the hands-off control is also called as {\em green control}~\cite{NagQueNes14a}.
Also, the hands-off control is effective in hybrid/electric vehicles,
railway vehicles,
networked/embedded systems, to name a few \cite{NagQueNes14b}.

Maximum hands-off control is related to {\em sparsity}, which is 
widely studied in compressed sensing \cite{EldKut}.
Sparsity is also applied to control problems such as
networked control \cite{NagQueOst14,KonGooSer14},
security of control systems \cite{FawTabDig14},
state estimation \cite{SanWakVin14},
to name a few.

A mathematical difficulty in the maximum hands-off control is that
the cost function,  which is defined by the $L^0$ measure (the support length of a function), 
is highly nonlinear; it is discontinuous and non-convex.
To solve this problem, a recent work \cite{NagQueNes13,NagQueNes14b}
has proposed to reduce the problem to  an $L^1$ optimal control problem, and 
shown the equivalence between the maximum hands-off (or $L^0$ optimal) 
control and the $L^1$ optimal control under the assumption of normality.

Motivated by this work, we investigate the value function in the maximum hands-off control.
The value function is defined as the optimal value of the cost function of the optimal control problem.
It is important to show the continuity of the value function with respect to the initial state;
if the value function is continuous, then the optimality property is less sensitive against
uncertainties in the initial state.
Also, the value function may be used as a Lyapunov function when the
optimal control is adapted to model predictive control,
and the continuity is necessary for the function to be a Lyapunov function
\cite{MorLee99}.
Although the $L^0$ measure in the maximum hands-off control
is discontinuous and non-convex,
we prove that the value function
is a continuous and strictly convex function of the initial state in the reachable set,
under an assumption on the controlled plant model.

The present paper expands on our recent conference contribution
\cite{IkeNag15} by rearranging the contents and
incorporating analysis of convexity of the value function.

The remainder of this paper is organized as follows:
In Section~\ref{sec:preliminaries},
we give mathematical preliminaries for our subsequent discussion.
In Section~\ref{sec:problem},  we review the problem of
maximum hands-off control.
Section~\ref{sec:continuity} investigates the continuity of the value function
in maximum hands-off control,
and Section~\ref{sec:convexity} discusses its convexity.
Section~\ref{sec:example} presents an example of
maximum hands-off control
to illustrate the properties of continuity and convexity.
In Section~\ref{sec:conclusion}, we offer concluding remarks.

\section{Mathematical Preliminaries}
\label{sec:preliminaries}
This section reviews basic definitions, facts, and notation that will be used
throughout the paper.

Let $n$ be a positive integer.
For a vector $x\in{\mathbb{R}}^n$
and a scalar $\varepsilon>0$, 
the {\em $\varepsilon$-neighborhood} of $x$ is defined by
${\mathcal B}(x,\varepsilon)\triangleq\{y\in\mathbb{R}^n: \|y-x\|<\varepsilon\}$, 
where $\|\cdot\|$ denotes the Euclidean norm in ${\mathbb{R}}^n$.
Let ${\mathcal X}$ be a subset of ${\mathbb{R}}^n$.
A point $x\in {\mathcal X}$ is called an {\em interior point} of ${\mathcal X}$ if there exists $\varepsilon>0$ 
such that ${\mathcal B}(x,\varepsilon)\subset {\mathcal X}$. 
The {\em interior} of ${\mathcal X}$ is the set of all interior points of ${\mathcal X}$, 
and we denote the interior of ${\mathcal X}$ by $\mathrm{int}{\mathcal X}$.
A set ${\mathcal X}$ is said to be {\em open} if ${\mathcal X}=\mathrm{int}{\mathcal X}$. 
For example, $\mathrm{int}{\mathcal X}$ is open for every subset ${\mathcal X}\subset\mathbb{R}^n$.
A point $x\in\mathbb{R}^n$ is called an {\em adherent point} of ${\mathcal X}$ 
if ${\mathcal B}(x,\varepsilon)\cap {\mathcal X} \neq\emptyset$ for every $\varepsilon>0$, 
and the {\em closure} of ${\mathcal X}$ is the set of all adherent points of ${\mathcal X}$.
A set ${\mathcal X}\subset{\mathbb{R}}^n$ is said to be {\em closed} if 
${\mathcal X}=\overline{{\mathcal X}}$, where $\overline{{\mathcal X}}$ is the closure of ${\mathcal X}$.
The {\em boundary} of ${\mathcal X}$ is the set of all points in the closure of ${\mathcal X}$, 
not belonging to the interior of ${\mathcal X}$, and we denote the boundary of ${\mathcal X}$ by 
$\partial {\mathcal X}$, i.e., $\partial {\mathcal X}= \overline{{\mathcal X}}-\mathrm{int}{\mathcal X}$, 
where $\mathcal{X}_1-\mathcal{X}_2$ is the set of all points which belong to the set ${\mathcal X}_1$ 
but not to the set ${\mathcal X}_2$. 
In particular, if ${\mathcal X}$ is closed, then $\partial {\mathcal X}= {\mathcal X} - \mathrm{int} {\mathcal X}$, 
since ${\mathcal X}=\overline{{\mathcal{X}}}$.
A set ${\mathcal X}\subset{\mathbb R}^n$ is said to be {\em convex} if,
for any $x,y\in{\mathcal X}$ and any $\lambda\in[0,1]$,
$(1-\lambda)x+\lambda y$ belongs to ${\mathcal{X}}$.


A real-valued function $f$ defined on $\mathbb{R}^n$ is said to be 
{\em upper semi-continuous} on $\mathbb{R}^n$ if for every $\alpha\in\mathbb{R}$ the set
$\{x\in \mathbb{R}^n: f(x)<\alpha\}$
is open, and $f$ is said to be {\em lower semi-continuous}
on $\mathbb{R}^n$ if for every $\alpha\in\mathbb{R}$ the set
$\{x\in \mathbb{R}^n: f(x)>\alpha\}$
is open. It is known that a function $f$ is continuous on $\mathbb{R}^n$ if and only if it is upper and lower semi-continuous on $\mathbb{R}^n$; see e.g., \cite[pp. 37]{Rud}.

A real-valued function $f$ defined on a convex set ${\mathcal{C}}\subset{\mathbb{R}^n}$ 
is said to be {\em convex} if 
\begin{equation}
 f\bigl((1-\lambda)x+\lambda y\bigr)\leq(1-\lambda)f(x)+\lambda f(y),
 \label{eq:convex}
\end{equation}
for all $x$, $y\in \mathcal{C}$ and all $\lambda\in(0,1)$, and $f$ is said to be {\em strictly convex}
if the inequality \eqref{eq:convex} holds strictly whenever $x$ and $y$ are distinct points and $\lambda\in(0,1)$.

Let $T>0$. For a continuous-time signal $u(t)$ over a time interval $[0, T]$, 
we define its {\em $L^1$ and $L^{\infty}$ norms} respectively by
\[\|u\|_{1}\triangleq\int_{0}^{T}|u(t)| dt, \quad \|u\|_{\infty}\triangleq\sup_{t\in[0, T]}|u(t)|.\]
We define the support set of $u$, denoted by $\mathrm{supp}(u)$, by
the closure of the set $\{t\in[0,T]: u(t)\neq0\}$.
The {\em $L^0$ norm} of a measurable function $u$ as
the length of its support, that is,
$\|u\|_{0} \triangleq m\bigl(\mathrm{supp}(u)\bigr)$,
where $m$ is the Lebesgue measure on ${\mathbb{R}}$.

\section{Maximum Hands-off Control Problem}
\label{sec:problem}
In this paper, we consider a linear time-invariant system represented by
\begin{equation}
\dot{x}(t)=Ax(t)+Bu(t), \quad t\geq 0,
\label{eq:S}
\end{equation}
where $x(t)\in{\mathbb{R}}^n$, $u(t)\in{\mathbb{R}}$, $A\in{\mathbb{R}}^{n\times n}$,
and $B\in{\mathbb R}^{n\times 1}$.
Throughout this paper, we assume the following:
\begin{assum}
 \label{assum:meta-normal}
The pair $(A,B)$ is controllable and the matrix $A$ is nonsingular.
\end{assum}
Let $T>0$ be the final time of control.
For the system \eqref{eq:S}, we call a control $u=\{u(t):t\in[0,T]\}\in L^1$ \emph{admissible} if it steers 
$x(t)$ from a given initial state $x(0)=\xi\in\mathbb{R}^n$ to the origin at time $T$ (i.e., $x(T)=0$),
and satisfies the magnitude constraint
$\|u\|_{\infty}\leq1$.
We denote by ${\mathcal U}(\xi)$ the set of all admissible controls for an initial state $\xi\in{\mathbb{R}}^n$,
that is,
\begin{equation}
 {\mathcal{U}}(\xi) \triangleq \biggl\{u\in L^1: \int_0^T e^{-As}Bu(s)ds=-\xi, \|u\|_\infty\leq 1\biggr\}.
 \label{eq:setU}
\end{equation}

The {\it maximum hands-off control} is the minimum $L^0$-norm (or the sparsest) control
among the admissible control inputs.
This control problem is formulated as follows.
\begin{prob}[Maximum hands-off control]
\label{prob:L0}
For a given initial state $\xi \in {\mathbb R}^n$,
find an admissible control $u\in{\mathcal U}(\xi)$
that minimizes $J(u) = \|u\|_0$.
\end{prob}

\begin{figure}[tb]
\centering
\includegraphics[width=\linewidth]{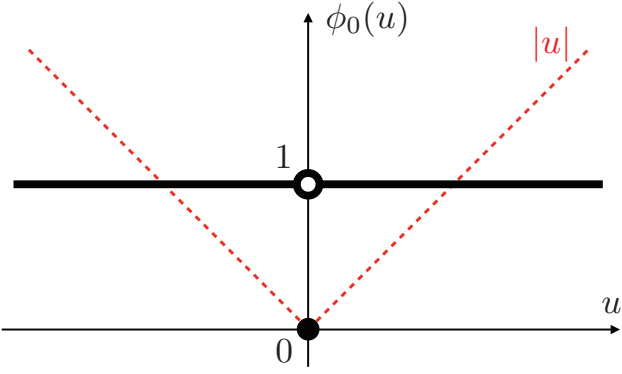}
\caption{The $L^0$ kernel $\phi_0(u)$ and its convex approximation $|u|$ for the $L^1$ norm.}
\label{fig:phi0}
\end{figure}

The value function for this optimal control problem is defined as
\begin{equation}
 V(\xi) \triangleq \min_{u\in{\mathcal U}(\xi)} J(u) = \min_{u\in{\mathcal U}(\xi)} \|u\|_0.
 \label{eq:Vx}
\end{equation}
Note that the cost function $J(u)$ can be rewritten as
\[
 J(u) = \int_0^T \phi_0(u)~dt,
\]
where $\phi_0$ is the $L^0$ kernel function defined by
\[
\phi_0(u) \triangleq \begin{cases} 1, & \text{~if~} u\neq 0,\\ 0, & \text{~if~} u=0. \end{cases}
\]
Fig.~\ref{fig:phi0} shows the graph of $\phi_0(u)$.
As shown in this figure, the kernel function $\phi_0(u)$ is discontinuous at $u=0$ and
non-convex.
However, in the following sections,
we will show that the value function $V(\xi)$ in \eqref{eq:Vx} is continuous and strictly convex.

\section{Continuity of Value Function}
\label{sec:continuity}
In this section, we investigate the continuity of the value function $V(\xi)$ in \eqref{eq:Vx}.

First, we define the {\em reachable set} for the control problem (Problem \ref{prob:L0})
by
\[
 {\mathcal{R}}\triangleq\bigg\{\int_{0}^{T}e^{-As}Bu(s) ds: \|u\|_{\infty}\leq1\bigg\}\subset{\mathbb R}^n.
\]
The following is a fundamental lemma of the paper:
\begin{lem}
\label{thm:equivalence}
Suppose Assumption \ref{assum:meta-normal} is satisfied.
Let us consider $L^1$ optimal control with
\begin{equation}
 \begin{split}
  J_1(u) &:= \|u\|_1 = \int_0^T |u(t)| dt,\\
  V_1(\xi) &:= \min_{u\in{\mathcal U}(\xi)} \|u\|_1.
 \end{split}
 \label{eq:J1}
\end{equation}
Then, for every $\xi\in{\mathcal{R}}$, we have
$V(\xi) = V_1(\xi)$.
\end{lem}
\begin{pf}
By Assumption \ref{assum:meta-normal}, 
the $L^1$-optimal control problem associate with \eqref{eq:J1}
is normal \cite[Theorem 6-13]{AthFal}.
Also, for $\xi\in{\mathcal{R}}$, an $L^1$-optimal control $u^\ast\in{\mathcal{U}}(\xi)$
minimizing $J_1$ exists (see Lemma \ref{lem:R_L1} in Appendix \ref{sec:appendix}), 
and $u^\ast(t)\in\{-1,0,1\}$ for almost all $t\in[0,T]$
(this is called the {\em ``bang-off-bang'' property}) \cite[Section 6-14]{AthFal}.
Then by \cite[Theorem 5]{NagQueNes13}, $u^\ast$ is also the optimal control of
Problem \ref{prob:L0}, and we have
\[
 V(\xi) = \min_{u\in{\mathcal U}(\xi)} \|u\|_0
  = \|u^\ast\|_0
  = \|u^\ast\|_1
  = V_1(\xi),
\]  
where we used the ``bang-off-bang'' property of $u^\ast$ for the third equality.
\qed
\end{pf}
Note that the absolute value $|u|$ in \eqref{eq:J1} is a convex approximation
of $\phi_0(u)$ as shown in Fig.~\ref{fig:phi0}.
Associated with $V_1(\xi)$, we define the following subset of ${\mathcal{R}}$
with $\alpha\geq 0$:
\begin{equation} 
 {\mathcal{R}}_{\alpha}\triangleq\bigg\{\int_{0}^{T}e^{-As}Bu(s) ds: \|u\|_{\infty}\leq1, \|u\|_{1}\leq\alpha\bigg\}.
 \label{eq:R_alpha_definition}
\end{equation} 
For the set ${\mathcal{R}}_{\alpha}$, we have another fundamental lemma.
\begin{lem}
\label{lem:R_alpha}
Suppose Assumption \ref{assum:meta-normal} is satisfied.
Then, for every $\alpha\in[0, T]$,
\begin{align}
{\mathcal{R}}_{\alpha}&=\{\xi\in {\mathcal{R}}:V(\xi)\leq \alpha\},
  \label{eq:lem:R_alpha_1}\\
\partial {\mathcal{R}}_{\alpha}&=\{\xi\in {\mathcal{R}}:V(\xi)=\alpha\},
  \label{eq:lem:R_alpha_2}\\
\mathrm{int}{\mathcal{R}}_{\alpha}&=\{\xi\in {\mathcal{R}}:V(\xi)<\alpha\}.
  \label{eq:lem:R_alpha_3}
\end{align}
\end{lem}
\begin{pf}
See Appendix \ref{sec:appendix}.
\qed
\end{pf}

From these lemmas, we show the continuity of the value function $V(\xi)$.
\begin{thm}
\label{thm:continuity}
If Assumption \ref{assum:meta-normal} is satisfied,
then $V(\xi)$ is continuous on ${\mathcal R}$.
\end{thm}
\begin{pf}
Define
\[
 \overline{V}(\xi)\triangleq\begin{cases}V(\xi),&\text{~if~}\xi\in {\mathcal{R}},\\T,&\text{~if~}\xi\in{\mathbb{R}}^n-{\mathcal{R}}.\end{cases}
\]
It is enough to show that $\overline{V}(\xi)$ is continuous on $\mathbb{R}^n$.

First, we show that the set
\begin{equation}
 \{\xi\in \mathbb{R}^n: \overline{V}(\xi)<\alpha\}
 \label{eq:levelset1}
\end{equation}
is open for every $\alpha\in\mathbb{R}$.
If $\alpha\leq0$, then the set \eqref{eq:levelset1} is empty since
for any $\xi\in{\mathbb R}^n$, $\overline{V}(\xi)\geq 0$.
If $\alpha>T$, then the set \eqref{eq:levelset1} is $\mathbb{R}^n$,
since for any $\xi\in{\mathcal{R}}$, $V(\xi)\leq T$.
If $0<\alpha\leq T$, then the set \eqref{eq:levelset1} is a subset of ${\mathcal{R}}$,
and coincides with $\mathrm{int}{\mathcal{R}}_{\alpha}$ 
by Lemma \ref{lem:R_alpha}. 
Therefore, the set \eqref{eq:levelset1} is open for every $\alpha\in\mathbb{R}$. 
It follows that $V(\xi)$ is upper semi-continuous on $\mathbb{R}^n$.

Next, we show that the set 
\begin{equation}
 \{\xi\in\mathbb{R}^n: \overline{V}(\xi)>\alpha\}
 \label{eq:levelset2}
\end{equation}
is open for every $\alpha\in\mathbb{R}$.
If $\alpha<0$ or $\alpha\geq T$, then the set \eqref{eq:levelset2}
is $\mathbb{R}^n$ or empty, respectively.
If $0\leq\alpha<T$, from Lemma \ref{lem:R_alpha}, we have
\[
 \begin{split}
  \{\xi\in\mathbb{R}^n: \overline{V}(\xi)>\alpha\}
    &=\mathbb{R}^n-\{\xi\in {\mathcal{R}}: V(\xi)\leq\alpha\}\\
    &=\mathbb{R}^n-{\mathcal{R}}_{\alpha}.
 \end{split}
\]
Since ${\mathcal{R}}_{\alpha}$ is closed (see Lemma \ref{lem:R_alpha_fundamental} in Appendix~\ref{sec:appendix}), 
the set \eqref{eq:levelset2} is open for every $\alpha\in\mathbb{R}$. 
It follows that $\overline{V}(\xi)$ is lower semi-continuous on $\mathbb{R}^n$.

Since $\overline{V}(\xi)$ is upper and lower semi-continuous on $\mathbb{R}^n$,
it is continuous on $\mathbb{R}^n$, and the conclusion follows.
\qed
\end{pf}

Theorem \ref{thm:continuity} leads to an important result of
$L^1$ optimal control as follows.
\begin{cor}
If Assumption \ref{assum:meta-normal} is satisfied,
then $V_1(\xi)$ is continuous on ${\mathcal R}$.
\end{cor}
\begin{pf}
This is a direct consequence of Lemma \ref{thm:equivalence} and Theorem \ref{thm:continuity}.
\qed
\end{pf}

\section{Convexity of Value Function}
\label{sec:convexity}
Here we show the convexity of the value function $V(\xi)$.
Although the kernel function $\phi_0(u)$ in the cost function is not 
convex as shown in Fig.~\ref{fig:phi0}, the value function
$V(\xi)$ is a convex function on ${\mathcal{R}}$.
\begin{thm}
If Assumption \ref{assum:meta-normal} is satisfied,
then $V(\xi)$ is strictly convex on ${\mathcal{R}}$.
\end{thm}
\begin{pf}
From Lemma \ref{thm:equivalence}, it is enough to prove that 
the $L^1$ value function $V_1(\xi)$ is strictly convex on ${\mathcal{R}}$

First, we prove that $V_1(\xi)$ is convex on ${\mathcal{R}}$.
Take any $\xi$, $\eta\in {\mathcal{R}}$, and $\lambda\in(0,1)$.
Then there exist $L^1$-optimal controls $u_{\xi}$ and $u_{\eta}$ for initial states $\xi$ and $\eta$, respectively
(see Lemma \ref{lem:R_L1} in Appendix \ref{sec:appendix}).
Obviously, the following control 
\begin{equation}
 u \triangleq (1-\lambda)u_{\xi}+\lambda u_{\eta}
 \label{eq:u}
\end{equation} 
steers the state from the initial state $(1-\lambda)\xi+\lambda \eta$ 
to the origin at time $T$, and it satisfies $\|u\|_{\infty}\leq1$.
That is,  we have $u\in {\mathcal{U}}\bigl((1-\lambda)\xi+\lambda \eta\bigr)$. 
Therefore
\begin{equation}
 \begin{split}
  V_1\bigl((1-\lambda)\xi+\lambda \eta\bigr)&\leq\|u\|_{1}\\
   &\leq (1-\lambda)\|u_{\xi}\|_{1}+\lambda\|u_{\eta}\|_{1}\\
   &=(1-\lambda)V_1(\xi)+\lambda V_1(\eta),
 \end{split}
 \label{eq:V1_convex}
\end{equation}
and hence $V_1(\xi)$ is convex on ${\mathcal{R}}$.

Next, we will show the strict convexity of $V(\xi)$.
To prove this, we will show that a contradiction is implied by assuming
that there exist $\xi$, $\eta\in {\mathcal{R}}$ with $\xi\neq\eta$ and $\lambda\in (0,1)$ such that 
\begin{equation}
 V_1\bigl((1-\lambda)\xi+\lambda \eta\bigr)=(1-\lambda)V_1(\xi)+\lambda V_1(\eta).
  \label{eq:V1_false}
\end{equation}
Let $u_{\xi}$ and $u_{\eta}$ be $L^1$-optimal controls 
for initial states $\xi$ and $\eta$, respectively.
Let
$u \triangleq (1-\lambda)u_{\xi}+\lambda u_{\eta}$
as in \eqref{eq:u}.
From \eqref{eq:V1_convex} and \eqref{eq:V1_false},
it follows that 
\[
 V_1\bigl((1-\lambda)\xi+\lambda \eta\bigr)=\|u\|_{1}=(1-\lambda)\|u_{\xi}\|_{1}+\lambda\|u_{\eta}\|_{1},
\]
so the control $u=(1-\lambda)u_{\xi}+\lambda u_{\eta}$ is an $L^1$-optimal control
for the initial state $(1-\lambda)\xi+\lambda \eta$.

Now, by Assumption \ref{assum:meta-normal},
$u_{\xi}(t)$ and $u_{\eta}(t)$ take the values $1$, $0$, and $-1$ at almost all $t\in[0,T]$.
So, the pair $(u_{\xi}(t), u_{\eta}(t))$ takes 
the following values on $[0, T]$ except for sets of measure zero:
\begin{equation}
\begin{split}
&(1, 1), (1, 0), (1, -1), (0, 1), (0, 0),\\
&(0, -1), (-1, 1), (-1, 0), (-1, -1).
\end{split}
\label{eq:pair}
\end{equation}
For the pairs in \eqref{eq:pair} of $(u_\xi(t),u_\eta(t))$,
the control $u=(1-\lambda)u_{\xi}+\lambda u_{\eta}$ respectively takes the following values:
\[
 1, 1-\lambda, 1-2\lambda, \lambda, 0, -\lambda, -1+2\lambda, -1+\lambda, -1.
\]
On the other hand,
the control $u$ is also $L^1$ optimal and takes the values $1$, $0$, and $-1$ at almost all $t\in[0, T]$.
Since $\lambda\in(0,1)$, we have
\begin{equation}
  m({\mathcal{I}}_{1,0}\cup{\mathcal{I}}_{0,1}\cup{\mathcal{I}}_{0,-1}\cup{\mathcal{I}}_{-1,0})=0,
  \label{eq:m_zero}
\end{equation}
where
${\mathcal I}_{i,j} \triangleq \{t\in[0,T]: (u_\xi(t),u_\eta(t)) = (i,j)\}$,
for $i,j\in\{-1,0,1\}$.
If $\lambda\neq 1/2$, then we also have
\[
 m({\mathcal{I}}_{1,-1}\cup{\mathcal{I}}_{-1,1})=0,
\] 
and it follows that 
\[
m({\mathcal{I}}_{1,1}\cup{\mathcal{I}}_{0,0}\cup{\mathcal{I}}_{-1,-1})=T,
\]
that is, $u_\xi(t)=u_\eta(t)$ for almost all $t\in[0,T]$.
This implies $\xi=\eta$, but this contradicts the assumption, so we have $\lambda=1/2$.
Then the pair $(u_{\xi}(t), u_{\eta}(t))$ on $[0, T]$ except for sets of measure zero
takes values $(1, 1)$, $(1, -1)$, $(0, 0)$, $(-1, 1)$, and $(-1, -1)$. 
Since $\xi\neq\eta$, we have
\begin{equation}
 T_1\triangleq m({\mathcal{I}}_{1,-1}\cup{\mathcal{I}}_{-1,1})>0.
 \label{eq:T1}
\end{equation} 
Let 
$T_2 \triangleq m({\mathcal{I}}_{1,1})$ and
$T_3 \triangleq m({\mathcal{I}}_{-1,-1})$.
From \eqref{eq:m_zero} and the fact that $u_\xi+u_\eta=0$ on ${\mathcal{I}}_{1,-1}\cup{\mathcal{I}}_{-1,1}\cup{\mathcal{I}}_{0,0}$, we have
\begin{equation}
 \begin{split}
  V_1\biggl(\frac{1}{2}\xi+\frac{1}{2}\eta\biggr)
   &=\biggl\|\frac{1}{2}u_{\xi}+\frac{1}{2}u_{\eta}\biggr\|_{1}\\
   &=\frac{1}{2}\int_{{\mathcal{I}}_{1,1}\cup{\mathcal{I}}_{-1,-1}} |u_\xi(t)+u_\eta(t)| dt\\
   &=T_{2}+T_{3},
 \end{split}
 \label{eq:V1_1}
\end{equation}
On the other hand,
\begin{equation}
 \begin{split}   
   \frac{1}{2}V_1(\xi)+\frac{1}{2}V_1(\eta) &= \frac{1}{2}\|u_\xi\|_1+\frac{1}{2}\|u_\eta\|_1\\
   &= T_1+T_2+T_3.
 \end{split}  
 \label{eq:V1_2}
\end{equation}
Equations \eqref{eq:V1_false}, \eqref{eq:V1_1} and \eqref{eq:V1_2}
imply that $T_1=0$, which contradicts \eqref{eq:T1}.
\qed
\end{pf}

\section{Example}
\label{sec:example}
In this section, we consider a simple example with a 1-dimensional linear control system 
\[
 \dot{x}(t)=ax(t)+bu(t),
\]
where $a<0$ and $b\neq0$.
This system
obviously satisfies Assumption \ref{assum:meta-normal},
and let us verify the continuity and convexity of 
the value function $V(\xi)$ on the reachable set ${\mathcal{R}}$.

The reachable set ${\mathcal{R}}$ and the maximum hands-off control $u_\xi$ for an initial state $\xi\in{\mathcal{R}}$ are computed via the bang-bang principle \cite[Theorem 12.1]{HerLas} and the minimum principle for $L^1$-optimal control \cite[Section 6.14]{AthFal} as
\[
 {\mathcal{R}}=[-x_{1}, x_{1}],\quad x_1=-|b|a^{-1}\bigl(e^{-aT}-1\bigr),
\] 
and 
\[
u_\xi(t)=\begin{cases}0,&t\in[0, \tau_\xi),\\-\mathrm{sgn}(b) \mathrm{sgn}(\xi),&t\in[\tau_\xi, T],\end{cases}
\]
where $\mathrm{sgn}(x)= x/|x|$ for $x\neq 0$ and $\mathrm{sgn}(0)=0$, and
\[
 \tau_\xi \triangleq -a^{-1}\log \bigl(e^{-aT}+a|b^{-1}\xi|\bigr).
\]
Note that if $\xi=0$, then $u_0(t)=0$ for all $t\in[0, T]$.
Then we have 
\[
V(\xi) = T-\tau_\xi=T+a^{-1}\log(e^{-aT}+a|b^{-1}\xi|).
\]

\begin{figure}[h]
\centering
   \includegraphics[width=\linewidth]{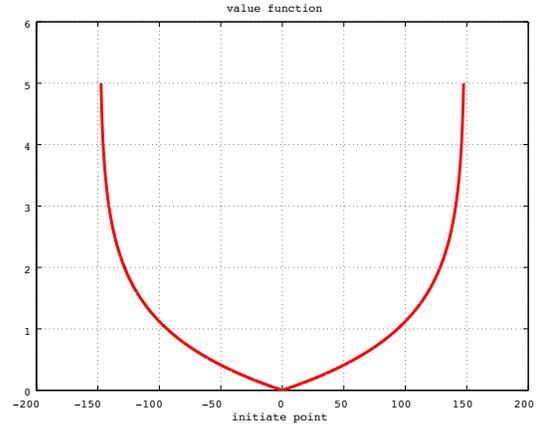}
  \caption{Value function $V(\xi)$ for $\xi\in {\mathcal{R}}=[-e^5+1,e^5-1]$}
  \label{fig:V}
\end{figure}

For example, let $a=-1$, $b=1$, and $T=5$. Fig.~\ref{fig:V} shows the value function $V(\xi)$ on ${\mathcal{R}}$, 
where ${\mathcal{R}}=[-e^{5}+1, e^{5}-1]$. Certainly, we can see that $V(\xi)$ is continuous 
and strictly convex on ${\mathcal{R}}$.

\section{Conclusion}
\label{sec:conclusion}
In this brief paper, we have proved the continuity and the strict convexity
of the value function of the maximum hands-off control
problem under an assumption of the controlled system.
Also, as a corollary we have shown that 
those properties are also satisfied for $L^1$ optimal control
under the same assumption.
These properties of the vale function plays an important role to investigate
the stability when we extend the control to the model predictive control. 

\begin{ack}                               
This research is supported in part by JSPS Grant-in-Aid for Scientific Research (C) No.~24560543, 
Grant-in-Aid for Scientific Research on Innovative Areas No.~26120521,
and an Okawa Foundation Research Grant.
\end{ack}

\bibliographystyle{plain}        
\bibliography{abrv,autosam}           



\appendix
\section{Proof of Lemma \ref{lem:R_alpha}}
\label{sec:appendix}
\subsection{Lemmas}
To prove Lemma \ref{lem:R_alpha}, we need some lemmas.
\begin{lem}
\label{lem:R_alpha_fundamental}
The set $R_\alpha$ in \eqref{eq:R_alpha_definition} satisfies the following:
\begin{enumerate}
\item For every $\alpha\in{\mathbb R}$, ${\mathcal{R}}_{\alpha}$ is compact.
\item For every $\alpha\in{\mathbb R}$, ${\mathcal{R}}_{\alpha}\subset {\mathcal{R}}$, with equality for $\alpha\geq T$.
\item ${\mathcal{R}}_0=\{0\}$.
\item ${\mathcal{R}}_{\alpha}\subset {\mathcal{R}}_{\beta}$ for $0\leq\alpha\leq\beta$.
\end{enumerate}
\end{lem}
\begin{pf}
See \cite[Lemma 2.1]{Haj79}.
\qed
\end{pf}
\begin{lem}
\label{lem:R_alpha_reachable}
For every $\alpha\in[0, T]$, we have
\[
 {\mathcal{R}}_\alpha=\{\xi\in {\mathcal{R}}:\exists\,u\in {\mathcal{U}}(\xi) {\rm~s.t.~} \|u\|_{1}\leq\alpha\}.
 \]
\end{lem}
\begin{pf}
First, fix $\alpha\in[0, T]$ and take any $\xi\in {\mathcal{R}}_{\alpha}$.
Then, by the definition of ${\mathcal{R}}_{\alpha}$,
there exists $u\in{\mathcal U}(\xi)$ such that
$\|u\|_1\leq \alpha$ and
\[
 \xi = \int_0^T e^{-As}Bu(s)ds.
\]
From \eqref{eq:setU}, it follows that
the control $v:=-u$ is an admissible control, that is, $v\in{\mathcal U}(\xi)$,
and also satisfies $\|v\|_1=\|u\|_1\leq \alpha$.
By definition, ${\mathcal{R}}_\alpha\subset  {\mathcal{R}}$ and hence $\xi\in{\mathcal{R}}$.
Therefore, we have $\xi\in \{\xi\in {\mathcal{R}}:\exists\, u\in {\mathcal{U}}(\xi) \text{~s.t.~} \|u\|_{1}\leq\alpha\}$.

Conversely, fix $\alpha\in [0, T]$ and take any
$\xi\in \{\xi\in {\mathcal{R}}:\exists\, u\in {\mathcal{U}}(\xi) \text{~s.t.~} \|u\|_{1}\leq\alpha\}$.
That is, $\xi\in {\mathcal{R}} $ is an initial state for the system \eqref{eq:S},
and there exists an admissible control $u\in {\mathcal{U}}(\xi)$ such that $\|u\|_{1}\leq\alpha$.
Then 
from \eqref{eq:setU}, we have
\[
 \xi=\int_{0}^{T} e^{-As}B\bigl(-u(s)\bigr)ds.
\]
The control $v=-u$ satisfies
$\|v\|_{1}=\|u\|_{1}\leq\alpha$,
$\|v\|_{\infty}=\|u\|_{\infty}\leq1$,
and hence
we have $\xi\in {\mathcal{R}}_{\alpha}$.
\qed
\end{pf}

\begin{lem}
\label{lem:R_L1}
For each initial value $\xi\in{\mathcal{R}}$, there exists
an admissible control $u\in{\mathcal{U}}(\xi)$ with minimal 
$L^1$-cost $\|u\|_1$.
Furthermore, then, $\xi\in\partial{\mathcal{R}}_\alpha$ with $\alpha=\|u\|_1$.
\end{lem}
\begin{pf}
See \cite[Lemma 3.1]{Haj79}.
\qed
\end{pf}

\subsection{Proof of \eqref{eq:lem:R_alpha_1}}
First, fix $\alpha\in [0, T]$ and take any $\xi \in {\mathcal{R}}_{\alpha}$.
Then, from Lemma \ref{lem:R_alpha_fundamental},
we have $\xi\in {\mathcal{R}}$, and from Lemma \ref{lem:R_L1},
there exists an $L^1$-optimal control $u^{\ast}\in{\mathcal{U}}(\xi)$.
Also, we have $V_1(\xi)=\|u^{\ast}\|_{1}\leq\alpha$ by Lemma \ref{lem:R_alpha_reachable}.
Then, from Lemma \ref{thm:equivalence}, we have $V(\xi)\leq\alpha$.
That is, we have
$\xi\in\{\xi\in {\mathcal{R}}:V(\xi)\leq \alpha\}$.

Conversely, fix $\alpha\in [0, T]$ and take any $\xi\in \{\xi\in {\mathcal{R}}:V(\xi)\leq \alpha\}$.
From Lemma \ref{thm:equivalence}, we have $V_1(\xi)\leq \alpha$.
Let $\beta\triangleq V_1(\xi)$. 
From Lemma \ref{lem:R_L1}, we have $\xi\in\partial {\mathcal{R}}_{\beta}$, 
and it follows from Lemma \ref{lem:R_alpha_fundamental}
that $\xi\in\partial {\mathcal{R}}_{\beta}\subset {\mathcal{R}}_{\beta}\subset {\mathcal{R}}_{\alpha}$.

\subsection{Proof of \eqref{eq:lem:R_alpha_2} and \eqref{eq:lem:R_alpha_3}}
We prove the equation \eqref{eq:lem:R_alpha_2};
then the equation \eqref{eq:lem:R_alpha_3} follows immediately from \eqref{eq:lem:R_alpha_1}
and \eqref{eq:lem:R_alpha_2}, since ${\mathcal{R}}_{\alpha}$ is closed for every $\alpha\geq0$
from Lemma \ref{lem:R_alpha_fundamental}.
If $\alpha=0$, then $\partial {\mathcal{R}}_{0}=\{0\}$, since ${\mathcal{R}}_{0}=\{0\}$. 
It follows from \eqref{eq:lem:R_alpha_1} that
\[\{\xi\in {\mathcal{R}}:V(\xi)=0\}={\mathcal{R}}_{0}=\{0\}=\partial {\mathcal{R}}_{0}.\]
Fix $\alpha\in(0, T]$. We can take $\xi\in\partial {\mathcal{R}}_{\alpha}$, 
since $\partial {\mathcal{R}}_{\alpha}$ is not empty.%
\footnote{$\mathbb{R}^n$ and the empty set are the only subsets 
whose boundaries are empty, since $\mathbb{R}^n$ is connected \cite[Chapter 3]{Sin}.}
Since $\xi\in {\mathcal{R}}_{\alpha}$, we have $V(\xi)\leq\alpha$. If $V(\xi)<\alpha$, 
then $\xi\in\partial {\mathcal{R}}_{V(\xi)}\subset {\mathcal{R}}_{V(\xi)}\subset$ int ${\mathcal{R}}_{\alpha}$
(see \cite[Lemma 4.2]{Haj79}),
and hence a contradiction occurs. 
Therefore we have $V(\xi)=\alpha$, 
and hence $\partial {\mathcal{R}}_{\alpha}\subset\{\xi\in {\mathcal{R}}:V(\xi)=\alpha\}$ 
and $\{\xi\in {\mathcal{R}}:V(\xi)=\alpha\}$ is not empty for every $\alpha\in(0, T]$.
Then it follows from Lemma \ref{lem:R_L1} that
\[\{\xi\in {\mathcal{R}}:V(\xi)=\alpha\}\subset\partial {\mathcal{R}}_{\alpha}\] 
for every $\alpha\in(0, T]$, and the conclusion follows.
\end{document}